\documentclass[preprint,12pt]{elsarticle}

\usepackage{graphicx}

\journal{Physica A}

\begin{document}

\begin{frontmatter}



\title{Phonon soft modes and para- to ferroelectric phase transitions}


\author{Jian-Sheng Wang}

\address{Department of Physics, National University of Singapore, Singapore 117551}

\begin{abstract}
In this special issue of Physica A  in memory of Professor Dietrich Stauffer, I first recall my impression on him while being
his postdoc for a year in HLRZ.   In the following scientific part, I discuss the theory of soft phonons with quartic nonlinear
interactions.  This is applied to the cubic crystal BaTiO$_3$ for phase transition from para- to ferroelectric phase.   
\end{abstract}


\end{frontmatter}

\section{Dedication to Professor Dietrich Stauffer}
\label{S:1}
Dietrich Stauffer was my postdoctoral supervisor for one year, starting March 1989.  As we know he passed away on 
6 August 2019 after a prolonged battle with cancer.  Over the past four decades he has made lasting contributions in very 
broad fields of science from the theory of phase transitions, percolation, and soft matter to bio-, econo- and sociophysics. 
In this first section, before I go into science proper, I like to have some reminiscence from my personal experience.

It was in the late 1980s,  after one postdoc work with Joel L. Lebowitz at Rutgers University, I was facing visa problem as J-1 visa allowed only so-called practical training period of eighteen months, so I had to find a job elsewhere outside US.
Professor Stauffer was already a famous scientist at the time due to his work on percolation and his well-known book on the subject.  My Ph.D. research subject in the cluster algorithms with Professor Robert Swendsen at Carnegie Mellon University 
is strongly influenced by the percolation simulation.   So, I was quite familiar with the subject matter and approached Stauffer for a postdoc position.   He was very forthcoming and offered me a job at HLRZ
(the full name is Höchstleistungsrechenzentrum, in Jülich, Germany).  This research institute was new and 
Stauffer was one of the two founding directors of HLRZ. He headed the statistical physics group. The other group at HLRZ that started simultaneously with the statistical physics group was on lattice gauge theory. 
Another postdoc of Stauffer, Subhrangshu Sekhar Manna, working on sand piles, jointed his group roughly about the same time, and we shared an office at the end of the hallway in a somewhat temporary two-storey building.  I remember Prof.~Stauffer had a lot of collaborators, so this also gave me advantage to collaborate with them.
I would like to mention my collaboration there with Debashish Chowdhury. An ex-postdoc of Stauffer and already young faculty member in Delhi, he came to HLRZ as a summer visitor when our collaboration began and then continued over the next two decades.
I collaborated with a few others through Stauffer, they are Walter Selke at KFA (Solid State Institute of Kernforschungsanlage, now is renamed Forschungszentrum Jülich, in which HLRZ is located), 
Ras B. Pandey (University of Mississippi), Volker Dohm (Aachen),  Heinz Mühlenbein (Gesellschaft fur Mathematik und Datenverarbeitung), Tane S. Ray (University of the West Indies), and few others from the USSR through the collaborators of the collaborators.   For a junior researcher in a year, this is a remarkable achievement, all because of Professor Stauffer's insistence on collaborations. 

Curiously, one may find that I have only one paper with Stauffer himself.  This paper was  the ``fractal dimension of 3D Ising droplets,'' published in Zeischrift Physik B in 1990 \cite{with-stauffer}.    It was a small paper dedicated to Professor Wilhelm Brenig on the occasion of his 60th birthday. The reason that I had only one paper with Stauffer, even though I published about ten in a year \cite{withchowd-J-Phys-france,wang-physica-A-161,with-selke1,with-selke2,wang-physica-A-164,with-muelenbein,with-andreichenko,with-ray}, was due to Stauffer's rule --- he has to run the codes if you want him to be a co-author.   As a result, he mediated a lot of collaborative efforts  without him being an author.  One particular piece of work on disordered 3D Ising model is worth record here.   Professor Mühlenbein's group had a self-built parallel super-computer and we like to try our program on it.   I remember we had to write in a very special parallel programming environment called `occam'.   I finished the programming and Professor Stauffer drove his car to St. Augustin.  I remember we spent a day there to set up the things.  Mühlenbein's student Ms Wöhlert helped to run the program.   
She reported back the result by emails.  This resulted a paper as J.-S. Wang, M. Wöhlert, H. Mühlenbein, and D. Chowdhury, ``The three-dimensional dilute Ising magnet,'' published in this exact journal, Physica A in 1990 \cite{with-muelenbein}.   

It is clear that Professor Stauffer has been dedicated to his service to the community, nurtured young minds 
like me at the time, helped scientists lacking computational resources to find their niche in research.  Due to special
nationality restriction, I was not allowed to use the Cray supercomputers hosted at KFA.  Prof.~Stauffer made special arrangement to 
allowed me to use the IBM 3090 mainframe frontend during the weekends for long runs. 

\section{Soft modes, an introduction}
After the above section of reminiscence of Professor Stauffer, I will present a current research topic, in the spirit of
Stauffer -- sort and concise.   This is a topic my group is working on (mostly by Dr.~Zhibin Gao) in connection with first principles calculations.   
It is about soft modes of phonons.  The soft phonon or self-consistent phonon concept is not new – this has a long history (Homer \cite{horner67}, Werthhamer \cite{werthhamer}, and others) and the Russian scientists developed  $\phi^4$-like field theory   (Larkin and Khmelnitskii \cite{larkin-JETP69,khmelnitskii73} and most recently from Coleman group \cite{palova09}) to study so-called displacement type phase transitions.  The soft phonons are clearly very relevant for para- to ferroelectric phase transitions \cite{cochran60,venkataraman79}.  Theory of such transitions is also not new; it is in Lines and Glass’s book on 
ferroelectrics \cite{lines-glass}.   The recent development up to 2006 is reviewed in Chapter 134 of Ghosez and Junquera \cite{ghosez06}.   That book chapter described the so-called Devonshire-Ginzburg Landau theory (1949) that is an even older, phenomenological theory.  The modern generalization is to fit the parameters to the first principle calculations they called effective Hamiltonian approach.  They call it “first principles”, but to me, it is still semi-phenomenological.

The non-interacting phonons are described by a quadratic form, with the Hamiltonian, in real space with discrete sites,
$H  = \frac{1}{2} \dot{u}^2 + \frac{1}{2} u^T K u$, here $u$ is a column vector of mass renormalized displacement,
$u_j = \sqrt{M_j}\, x_j$, and $K$ is the force constant matrix.   The coupled equations of motion can be diagonalized and
the displacement variables expressed in terms of normal mode coordinates.  The quanta of the resulting independent harmonic oscillators are known as phonons.   The high order terms in a Taylor expansion of the potential can be viewed
as introducing phonon-phonon interactions.  For example, the most important phonon-phonon scattering progress is due to
the third order terms.   However, the fourth order terms are also important, and in a sense are simpler than the 3rd order
terms.   Here in this paper we focus only on the 4-th order nonlinear terms which is the subject of soft modes. 

The simplicity of the fourth-order term comes from the fact we can make a mean-field approximation by replacing $u^4$ by
$\langle u^2 \rangle u^2$, rendering a nonlinear problem back to a simpler quadratic problem with an effective force
constant that is determined self-consistently.   All of the soft phonon theories are more or less based on such ideas.   It is not clear at all that such a theory is accurate.   However, our concrete calculation of thermal transport on few
degree problems illustrates that it can be rather accurate \cite{second-review,he16}.    

\section{Feynman-Jensen-Bogoliubov inequality}
However, the above replacement, aside being missing a factor of 3 due to combinatorial factors, also missed out the possibility of the system develops ferroelectric order (appearance of dipole moments for ionic systems).   A slightly more
fundamental approach is to base on the Feynman-Jensen-Bogoliubov (FJB) inequality of the free energy.   Here we consider the following quantum-mechanical Hamiltonian,
\begin{equation}
\hat{H} = \frac{1}{2} p^2 + \frac{1}{2} u^T K u - u^T f + \frac{1}{4} \sum_{ijkl} T_{ijkl}\, u_i u_j u_k u_l,
\end{equation}
here $p = \dot{u}$ is the momentum conjugate to $u$, and the superscript $T$ stands for matrix transpose, and
$f$ is an external force that perturbs the system.  The last term is the fourth order non-linear interaction.  

We can develop the nonequilibrium Green's function (NEGF) formalism, but it will be faster if we only focus on the 
Helmholtz free energy,  $F = - k_B T \ln Z$.  According to the FJB inequality \cite{feynman72}
\begin{equation}
F \leq F_0 + \langle \hat{H} - \hat{H}_0 \rangle = \psi.
\end{equation}
Here we need to choose a reference system $\hat{H}_0$ which we take it to be 
$\hat{H}_0 = \frac{1}{2} \Delta \dot{u}^2 + \frac{1}{2} \Delta u^T K^{e} \Delta u$, which is a quadratic form
with $K^e$ and $\bar{u} = u  - \Delta u$ as adjustable parameters.   It is important that we measure the deviation 
from the average, particularly when we have a nonzero $f$.  The average $\langle \,\cdot\, \rangle$ is calculated in the
reference system.  The usual derivation is to set $\bar{u}$ to zero which will not be
able to describe the ordered states.  When the right-hand side of the inequality is 
evaluated, we minimize it and obtain a pair of equations determining these parameters, which will be our best estimate to
the true free energy.   Since this is a quantum system, the free energy can be expressed in terms of the
Matsubara Green's function $G$, we obtain, for the reference system ($\beta = 1/(k_B T)$),
\begin{equation}
F_0 = - \frac{1}{\beta} \ln Z_0 = \frac{1}{2\beta} {\rm Tr} \ln \bigl( - G^{-1}\bigr) + {\rm const}. 
\end{equation}
This equation is somewhat formal as a limiting process is needed for the sum of the Matsubara frequencies to converge,
but this problem goes away when we only need the variation of the above expression.

Since the reference system is a Gaussian form, Wick's theorem applies, but only to the deviation $\Delta u$.   After writing
$u$ as $\bar{u} + \Delta u$, we can then apply Wick's theorem, such as
\begin{eqnarray}
\langle \Delta u_i \Delta u_j \Delta u_k \Delta u_l \rangle &=& 
\langle \Delta u_i \Delta u_j \rangle \langle \Delta u_k \Delta u_l \rangle + \nonumber \\ 
&& \langle \Delta u_i \Delta u_k \rangle \langle \Delta u_j \Delta u_l \rangle + 
\langle \Delta u_i \Delta u_l \rangle \langle \Delta u_j \Delta u_k \rangle.
\end{eqnarray}
Since there is a unique relation between the Green's function and the effective force constant through
$G(i\omega_n) = \bigl[ (i\omega_n)^2 - K^e\bigr]^{-1}$ here $\omega_n$ is the Matsubara frequencies, a variation in
$K^e$ can also be expressed in terms of variation in the Green's function $G$, given $\delta K^e = G^{-1} \delta G G^{-1}$.
The explicit expression for $\psi$ and its variation are messy but straightforward; I only give the final equations that minimize 
the free energy, which are
\begin{eqnarray}
\label{eqKe}
K^e &=& K + \Sigma_{\rm loop} + \Sigma_{\rm bunny}, \\
\label{equ}
f &=&  \left(K +  \Sigma_{\rm loop} + \frac{1}{3} \Sigma_{\rm bunny}\right) \bar u.
\end{eqnarray}
The first equation gives the renormalized force constant $K^e$ and second equation describes the force balance due to
external force $f$.  Notice the coefficient for the bunny-diagram term is different.   $\bar u$ determines the displacement.  A nonzero $\bar u$ indicates some sort of ordering.  Here we define the two ``self-energies'' as 
\begin{equation}
 \Sigma_{\rm loop}^{ij}  = 3 \sum_{kl} T_{ijkl} \langle \Delta u_k \Delta u_l \rangle,
\quad \Sigma_{\rm bunny}^{ij} = 3  \sum_{kl} T_{ijkl} \bar u_k \bar u_l.
\end{equation}
They are named so because the Feynman diagrams associated with these two terms.  The pair of equations need to be solved on computer self-consistently.

\section{Implementation and results}
Since we are interested in crystal lattices with lattice periodicity, it is best that the equations are transformed in mode space
such that $K$ is diagonal with eigen frequencies $\{ \omega_n({\bf q})^2 \}$.   In mode space, the displacement $u$ becomes the normal mode coordinate $Q$, so the pair of equations can be rewritten as 
\begin{eqnarray}
D_{nn'}({\bf q}) &= &   \omega_n({\bf q})^2  \delta_{nn'} + 
3 \sum_{mm'{\bf q}'} T_{nn'mm'}({\bf q}, {\bf q'}) \bigl\langle Q_m({\bf q}') Q_{m'}({\bf q}')^{*} \bigr\rangle  \nonumber  \\
\label{Dq} &&\quad +\, 3 \sum_{mm'} T_{nn'mm'}({\bf q}, {\bf 0}) \langle Q_m  \rangle \langle Q_{m'}^{*} \rangle, \\
\label{bunnyf}
f_n &=& \sum_{m} \left( D_{nm}({\bf 0})  - \frac{2}{3} \Sigma_{{\rm bunny}}^{nm} \right) \langle Q_m\rangle
\end{eqnarray}
Here the last equation is evaluated at ${\bf q}={\bf 0}$.   The matrix $D$ can be diagonalized  with the tilded version of frequency squares, $D = C \{ \tilde{\omega}^2_n \} C^\dagger$.
From this diagonal representation, we can evaluate the normal mode correlation for each $\bf q$ by
$\langle Q Q^\dagger \rangle = C \{ \frac{\hbar}{2\ \tilde{\omega}_n}( 2 N(\tilde\omega_n) +1) \} C^\dagger$, where
$N(\tilde\omega_n)=1/(e^{\beta \hbar \tilde \omega_n}-1)$ is the Bose function.  The curly braces denote diagonal matrices in the mode space with the
diagonal elements indicated. 

To calculate the dielectric response, we apply a small, uniform electric field $\bf E$ and compute the displacements of the atoms, thus the polarization.   The force generated by the electric field needs to be projected into the mode space, which is
\begin{equation}
f_n = e \sum_{k\alpha\beta} e_{k\alpha}^*(n) \frac{Z_{\beta \alpha}^k}{\sqrt{M_k}} E_\beta, 
\end{equation}
and the polarization is
\begin{equation}
P_\alpha = \frac{e}{\Omega} \sum_{k\beta n} \frac{Z^k_{\alpha\beta}}{\sqrt{M_k}} e_{k\beta}(n)
\langle Q_n \rangle.
\end{equation}
Here $e$ is the unit charge, $Z^k$ is the Born effective charge matrix of atom $k$, $e_{k\beta}(n)$ is the polarization vector of phonon mode $n$ (defined by $K$), $Q_n$ is the normal mode coordinate, both at the $\Gamma$ point. $\Omega$ is the volume of the unit cell.  From this we can deduce the dielectric susceptibility by $P_\alpha = \epsilon_0 \sum_{\beta} \chi_{\alpha\beta} E_\beta$, and the total relative dielectric constant is $\epsilon_{\rm tot} = \epsilon_\infty + \chi$, where $\epsilon_\infty$ is the electron contribution at low frequencies.  Above the critical temperature, the bunny diagram term
in Eq.~(\ref{bunnyf}) can be neglected since it is $O(\bar{u}^3)$, $\langle Q_n \rangle \approx f_n/\tilde\omega_n^2$
in the mode space defined by $K^e$. A more explicit expression can then be given for $\chi$, which agrees with \cite{bornhuang54,gonze97}
but with a renormalized or effective frequency.

\begin{figure}[t]
\centering\includegraphics[width=0.6\linewidth]{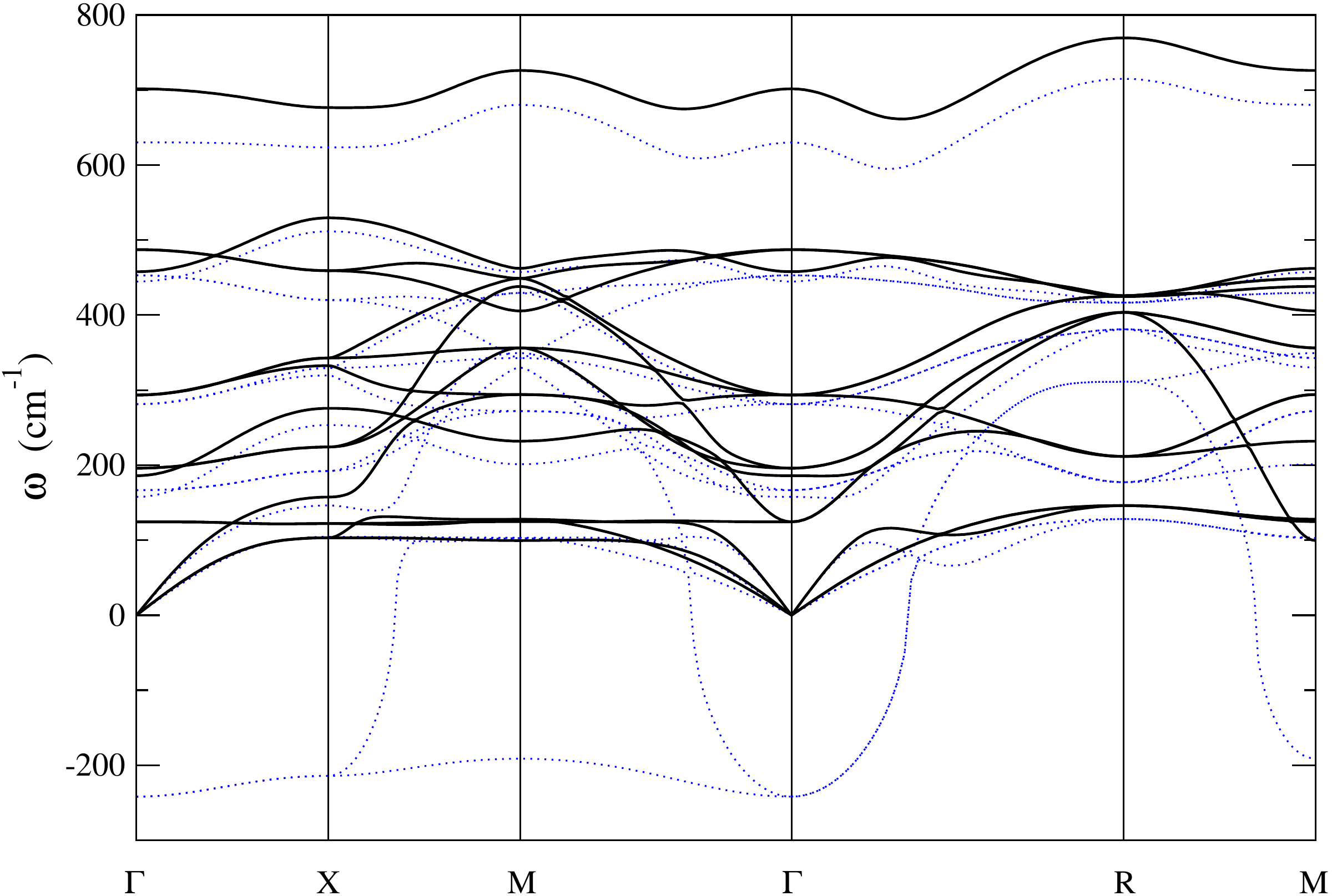}
\caption{The dispersion relation calculated with harmonic force constants (dotted lines) and with 4-th order 
nonlinear renormalized effective phonons (solid lines) at 400\,K.  The $\bf q$-point in Eq.(\ref{Dq}) is sampled on a 
$2\times 2 \times 2$ grids while ${\bf q}'$ on a denser grid of $8\times 8\times 8$.}
\end{figure}

As an example of the application of the theory, we apply to BaTiO$_3$.   This is one of the well-known systems that experience ferroelectric order. The crystal has a series of structure phase transitions from cubic, tetragonal, orthorhombic, to rhombohedral.  Here we focus only on the cubic and possibly tetragonal
phases at high temperatures.  
We run Quantum Espresso to determine the forces on a given ionic configuration with  supercell of $2\times 2\times 2$ of 40 atoms, with an energy cut of 60\,Ry and $4 \times 4 \times 4$ k-point sampling, using the PBE functional with projector augmented-wave pseudopotential.    The optimal lattice constant is found to be 4.026\,\AA. 
We determine the force constants following \cite{zhouPRL14,tadano15} using Tadano's codes
known as ALAMODE.   For the nonlinear force constants, we only go to 4-th order, with a distance cut-off of 12 bohr, 
 ignoring the numerous 4-body interactions.    The LASSO method identified about 3000 unique free parameters.    By
 looking at the fluctuations with different displacement-force data sets, we probably can determine the 4-th order
 force constants accurate to about 2 digits only. 
 
The renormalized phonon dispersion appears well-behaved, and all the imaginary frequencies can be made positive.  
We notice a flat band which has the lowest nonzero frequency at $\Gamma$ point.   This band corresponds to
a soft mode in the sense that it shifts down to lower values and eventually to 0, causing a large dielectric 
response and a phase transition, as envisaged by Cochran  \cite{cochran60} many years ago.   Unfortunately to
this naive author, the situation for BaTiO$_3$ is much more complicated than initially thought.   First, 
the predicted transition is much
too low in temperature comparing to the experimental value \cite{merz49} of about 400\,K (applying a negative pressure could raise the transition temperature \cite{seo13}).   Second, positions of the Ti atoms are 
known to shift to one of 8 equivalent $\langle 1,1,1\rangle$ directions, producing large entropic effect (the so-called order-disorder
transition \cite{liu16}) which the present theory apparently cannot capture.  According to Ref.~\cite{goddard-III06}, even in 
cubic phase the system also shows anti-ferroelectric character.   If so, our primitive cell of 5 atoms is too small 
to incorporate such an order.    I will try to identify better, purely displacive transition systems for which our theory is applicable, as a future research project.  These may be also useful in relation to the phonon Hall effect for which we are also very interested \cite{kangtai20}.

\begin{figure}[t]
\centering\includegraphics[width=0.7\linewidth]{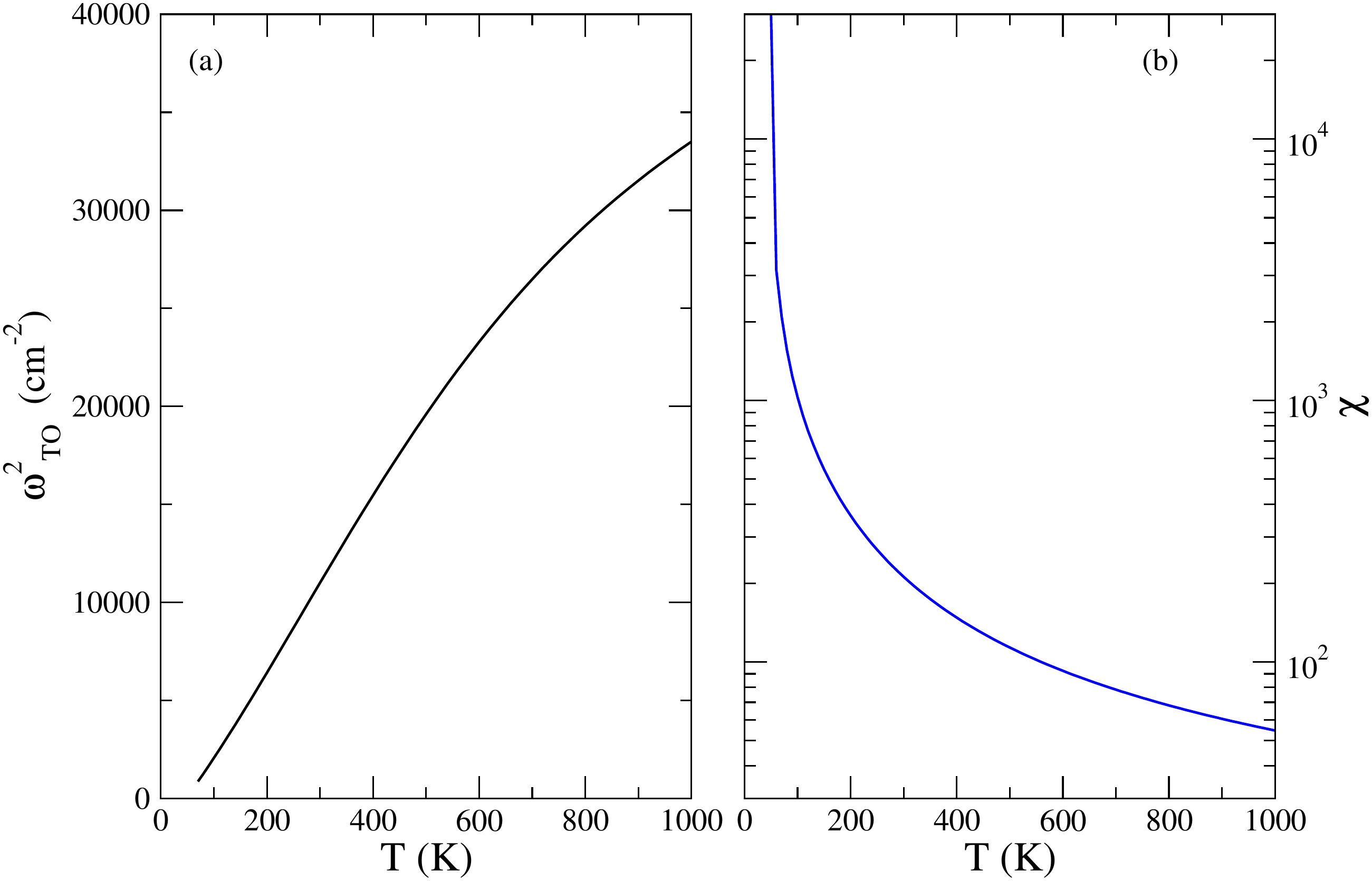}
\caption{(a) The renormalized soft mode frequency squared (the lowest transverse optical mode) as a function of
temperature $T$, using the same parameters as for the dispersion relation calculation.  (b) The dielectric susceptibility $\chi$ (ionic contribution to dielectric constant) as a function of temperature $T$.}
\end{figure}

The author thanks Prof.~Debashish Chowdhury for the invitation for a contribution to this special issue in memory 
of Professor Stauffer.  He thanks Dr. Zhibin Gao for drawing attention to the soft phonon problem and also teaching the
author on how to run first principles codes.  Without his help, much of the calculations would not be possible. This work was supported by a FRC grant R-144-000-402-114 and MOE tier 2 grant R-144-000-411-112.





\bibliographystyle{elsarticle-num}
\bibliography{stauffer}







\end{document}